\documentstyle[12pt]{article}

\author{E.~L. Afraimovich, E. I. Astafieva,\\
Institute of Solar-Terrestrial Physics SD RAS,\\
p.~o.~box~4026, Irkutsk, 664033, Russia,\\
fax: +7 3952 462557; e-mail:~afra@iszf.irk.ru}
\title{\large
{\vspace{3ex}
\bf Isolated ionospheric disturbances as deduced from
global GPS network}}
\date{}
\begin{document}
\maketitle
\begin{abstract}
We investigate an unusual class of medium-scale traveling
ionospheric disturbances (MS TIDs) of the nonwave type, isolated
ionospheric disturbances (IIDs) that manifest themselves in total
electron content (TEC) variations in the form of single aperiodic
negative TEC disturbances of a duration of about 10 min (the
total electron content spikes, TECS). For the first time, we
present the TECS morphology for 170 days with a different level
of geomagnetic activity and with the number of stations of the
global GPS network ranging from 4 to 240. A total number of the
TEC series (radio paths) used in the analysis, corresponding to
the observation along a single receiver-satellite Line-of-Sight
(LOS), with a duration of each series of about 2.3 hours,
exceeded 850000. The data were obtained using the technology of
global detection and monitoring of ionospheric disturbances
(GLOBDET, developed at the ISTP SB RAS) of a natural and
technogenic origin using measurements of TEC variations from a
global network of receivers of the GPS. It was found that TECS
are observed in no more than 1-2$\%$ of the total number of radio
paths. We present the results derived from analyzing the
dependence of TECS parameters on the local time, and on the level
of geomagnetic activity. The TECS amplitude exceeds at least one
order of magnitude the TEC fluctuation intensity under
"background" conditions. The IID-induced TEC variations are
similar in their amplitude, form and duration to the TEC response
to shock-acoustic waves (SAW) generated during rocket launchings
and earthquakes. However, the IID propagation velocity is less
than the SAW velocity (800-1000 m/s) and are most likely to
correspond to the velocity of background medium-scale
acoustic-gravity waves (AGW), on the order of 100-200 m/s.
\end{abstract}
\section*{Keywords}
Ionosphere (ionospheric irregularities, instruments and techniques), 
radio science (ionospheric propagation)

\section{Introduction}
\label{SPE-sect-1}
Of the known ionospheric irregularities of a different class,
mid-latitude isolated ionospheric disturbances (IIDs) stand out
as a highly unusual type. The past 40 years saw a consistent interest
in the study of the origin of IIDs that was aroused due to
difficulties in determining an adequate mechanism for IID generation
in mid-latitudes, as well as by the fact that the IID have a marked
effect on amplitude and phase characteristics of transionospheric
signals from radio engineering communication and navigation
systems (Afraimovich et al., 1992) by causing serious malfunctions
of these systems.

A large number of publications (e.g., Karasawa, 1985; Titheridge,
1971), including a review by Bowman (1989) were devoted to the
study along this line. IIDs are detected when recording amplitude
and phase scintillations of transionospheric radio signals in the
form of rarely occurring single aperiodic negative impulses with
a duration from a few to several tens of seconds (Karasawa,
1985). To name this uncommon type of scintillation Karasawa
(1985) seems to be the first to coin the term "spikes-type"
(S-) scintillations.

Karasawa et al. (1985) noticed from a long-term recording of the
signal from the geostationary MARISAT satellite at 1.5 GHz
frequency that synchronous with amplitude S-scintillations, there
occur similar-appearing changes of the rotation angle 
of polarization plane that are proportional to a corresponding
disturbance of total electron content (TEC). Anomalous
fluctuations, recorded during 13 months of observation, occur
predominantly in the night-time and last from 5 s to 2 min. The
diurnal dependence of the S-type oscillations shows two distinct
peaks: 09:00-15:00 in the daytime, and 20:00-01:00 at night. As
far as the seasonal dependence is concerned, however, a maximum
of the oscillation distribution corresponds to early summer
(June). It is customary to associate the occurrence of S-type
oscillations with diffraction or interference from small-scale
irregularities, "blobs" and "bubbles", generated in the
ionosphere.

Titheridge (1971) found that amplitude and phase
S-scintillations are caused by refraction and diffraction effects
at the propagation of the transionospheric signal in a medium
with IIDs and presented the corresponding formulae for estimating
these effects as a function of relationship of the wavelength of
the radio wave, the irregularity size and the sounding geometry
(LOS to the satellite and the distance to the layer with IIDs).

However, in spite of the many years of experimental and theoretical
investigations, there is as yet no clear understanding not only
of the physical nature of IIDs but even of their morphology (the
occurrence frequency as a function of geographical position,
time, level of geomagnetic and meteorological activity, etc.).

To tackle these questions requires statistically significant sets
of experimental data with good spatial and temporal resolution in
order to gain insight into not only morphological but also
dynamic IID characteristics: the direction of travel, the
propagation velocity, and the location of the possible
disturbance source. Another important requirement implies a
continuity and global coverage of observations since such
phenomena are relatively rare in time and random in space.

Such an opportunity is for the first time provided by the use of
the international ground-based network of two-frequency receivers
of the GPS, consisting of no less than 1000
sites as of the beginning of 2002 and posting its data on the
Internet, which open up new avenues for a global, continuous,
fully computerized monitoring of ionospheric disturbances of a
different class. Analysis and detection of IIDs was made possible
through the use of the technology for global detection and
determination of parameters of ionospheric disturbances of a
different class that was developed at the ISTP (Afraimovich, 2000b).

The objective of this paper is to study the morphology and
spatial and temporal properties of IIDs using data from a global
network of GPS receivers. Following Karasawa et al. (1985), the
term TECS (total electron content spikes) will be used here to
designate the IID-induced TEC disturbances. The sample statistic
of the occurrence frequency and morphology used in this study
does refer to TECS recorded from GPS data. Within the framework
of certain model representations, using these data it is possible
to reconstruct amplitude and spatial characteristics of local
electron density disturbances, i.e. of IIDs themselves. On this
basis, the term IIDs will be used below interchangeably with the
term TECS.

Section 2 describes the method for detecting the TECS obtained in
our study. Section 3 presents the TECS morphology. Section 4 is
devoted to a detailed analysis of the spatial and temporal
properties of IIDs by considering the most pronounced
manifestation of TECS on October 5, 2001 in California, USA. The
discussion of results compared with findings reported by other
authors is presented in Section 5.

Our comparison of IID characteristics with geomagnetic field
variations used data from near-lying magnetic variation stations
of the INTERMAGNET network (address: http://www.intermagnet.org).

\section{Method of processing the data from the global network.
            Selection of TECS}
\label{SPE-sect-2}

The standart GPS technology provides a means for wave
disturbances detecion based on phase measurements of TEC at each
of spaced two-frequency GPS receivers (Hofmann-Wellenhof et. al,
1992):

\begin{equation}
\label{IID-eq-01}
             I_o=\frac{1}{40{.}308}\frac{f^2_1f^2_2}{f^2_1-f^2_2}
                           [(L_1\lambda_1-L_2\lambda_2)+const+nL],
\end{equation}

where $L_1\lambda_1$ and $L_2\lambda_2$ are additional paths of
the radio signal caused by the phase delay in the
ionosphere,~(m); $L_1$ and $L_2$ represent the number of phase
rotations at the frequencies $f_1$ and $f_2$; $\lambda_1$ and
$\lambda_2$ stand for the corresponding wavelengths,~(m); $const$
is the unknown initial phase ambiguity,~(m); and $nL$~ are errors
in determining the phase path,~(m).

Phase measurements in the GPS can be made with a high degree of
accuracy corresponding to the error of TEC determination of at
least $10^{14}$~m${}^{-2}$ when averaged on a 30-second time
interval, with some uncertainty of the initial value of TEC,
however (Hofmann-Wellenhof et. al, 1992). This makes possible
detecting ionization
irregularities and wave processes in the ionosphere over a wide
range of amplitudes (up to $10^{-4}$ of the diurnal TEC
variation) and periods (from 24 hours to 5 min). The unit of TEC
$TECU$, which is equal to $10^{16}$~m${}^{-2}$~ and is commonly
accepted in the literature, will be used in the following.

Primary data include series of "oblique" values of TEC $I_0(t)$, as
well as the corresponding series of elevations $\theta_s(t)$ and
azimuths $\alpha_s(t)$ of the LOS to the satellite calculated using
our developed CONVTEC program which converts the GPS system
standard RINEX-files on the INTERNET (Gurtner, 1993).

Series of the values of elevations $\theta_s(t)$ and azimuths
$\alpha_s(t)$ of the LOS to the satellite were used to determine
the coordinates of subionospheric points, and to convert the
"oblique" TEC $I_{0}(t)$ to the corresponding value of the
"vertical" TEC by employing the technique reported by Klobuchar
(1986)

\begin{equation}
\label{IID-eq-02} I = I_0 \times cos
\left[arcsin\left(\frac{R_z}{R_z + h_{max}}cos\theta_s\right)
\right],
\end{equation}

where $R_{z}$ is the Earth's radius, and $h_{max}$=300 km is the
height of the $F_{2}$-layer maximum. All results in this study
were obtained for elevations $\theta_s(t)$ larger than
30$^\circ$.

The technology of global detection of TEC disturbances that was
developed at the ISTP SB RAS makes it possible to select - in the
automatic mode from an extensive amount of experimental material
- TEC disturbances which can be classed as TECS. TECS were
selected by two criteria. TEC variations were selected first, the
standard deviation (rms) of which exceeded the prescribed level
$\epsilon$. The statistic of TECS discussed in this paper was
obtained for $\epsilon$ = 0.1 $TECU$. Then, for each filtered
series we verified the fulfilment of the "singleness" condition
of the TEC overshoot.

Fig. 1 illustrates the TECS selection procedure. Fig. $1{\bf a}$
presents an example of weakly disturbed disturbances of a "vertical"
TEC $I(t)$ recorded on February 11, 2001 at station COCO ($96.8^\circ$E;
$12.1^\circ$S; the satellite number is PRN30). Fig. $1\bf b$ presents
the filtered (from the initial $I(t)$-series) $dI(t)$-variations. Thin
horizontal lines show the prescribed threshold $\epsilon$. The
standard deviation of the $dI(t)$-variations is 0.007 $TECU$, that is,
does not reach the prescribed threshold, $\epsilon$ = 0.1 $TECU$.

Figs. $1{\bf d}$ and $1{\bf e}$ plot the same dependencies as in
Figs. $1{\bf a}$ and $1{\bf b}$, but for station BAKO
($106.8^\circ$E; $06.5^\circ$S; the satellite number is PRN26).
It is evident from Fig. $1{\bf d}$ that unusual (for background
disturbances) TEC variations in the form of a single impulse of a
duration on the order of $\Delta T$=10 min are clearly identified
at the background of slow TEC variations.

The time $t_{min}$, corresponding to the value of $A_{min}$ of
the filtered TEC variation $dI(t)$, is shown in Fig.1{\bf e} by
the shaded triangle. The value of $\Delta T$ is determined from
the level of 0.5 $A_{min}$. The amplitude of this pulse $A_{min}$
far exceeds the specified threshold $\epsilon$ and at least an
order of magnitude higher than the background TEC fluctuation
intensity for this range of periods (Afraimovich et al., 2001b).

The relative amplitude of such a response $A_{min}/{I_0}$ has a
significant value, 2$\%$. We used, as the background value of
$I_0$, the absolute "vertical" TEC value $I_{0}(t)$ for the site
located at $7.5^\circ$S; $105^\circ$E, obtained from IONEX-maps
of TEC (Mannucci, 1998).

It should be noted that the above examples both refer to the same
time interval and to the stations spaced by a distance over
1300 km. This indicates a local character of the phenomenon and
is in agreement with the overall statistic characterizing its
spatial correlation (see Section 3).

For each of the events satisfying the above TECS selection
criteria, a special file was used to store information about the
GPS station name and geographic latitude and longitude; GPS
satellite PRN number; amplitude $A_{min}$; time $t_{min}$
corresponding to the minimum value of the $A_{min}$ amplitude;
and about the TECS duration $\Delta T$. The sample statistic,
presented below, was obtained by processing such files.

\section{Morphology of TECS}
\label{SPE-sect-3}

A total number of the TEC series (radio paths) used in the
analysis, corresponding to the observation along a single
receiver-satellite LOS, with a duration of each series of about
2.3 hours, exceeded 850000. The method outlined above was used to
obtain a set of TECS totaling about 10000 cases, or making up
about 1$\%$ of the total number of the LOS's. An analysis of the
resulting statistic revealed a number of dependencies of TECS
parameters on different factors.

First we consider the seasonal dependence of the TECS occurrence
frequency and amplitude (Fig. 2). Fig. $2{\bf a}$ plots the
number of days of observation $M$ versus season. As is evident,
the autumn is represented best statistically. Fig.$2{\bf b}$
shows the seasonal dependence of the number of TECS, $N$. Fig.
$2{\bf c}$ plots the number of TECS per day as a function of
season ${L = N/M}$. This dependence has maxima in spring and in
autumn.

The relative density of TECS, $D$, obtained as the ratio of the
number of TECS identified, $N$, to the number of LOS's, is
presented in Fig. $2{\bf d}$.

Vertical lines in Fig. $2{\bf d}$ show TECS standard deviations
(rms). The dashed horizontal line plots the threshold in amplitude
$\epsilon$ = 0.1 $TECU$. The most probable value of $<|A_{min}|>$
with a small scatter varies around the value 0.3 $TECU$,
independently of the season.

Fig. $3{\bf a}$ plots the dependence $P(|\it Dst|)$ of the number
of TECS on the modulus of values of the geomagnetic activity
index $\it Dst$. There is a general tendency of the number of
TECS to increase with the decreasing level of geomagnetic
activity. Most (80$\%$) of TECS occur when values of the $\it
Dst$-index are below 50 nT.

Fig. $3{\bf b}$ presents the diurnal distribution of TECS,
$P(t_{min})$, for the times $t_{min}$ corresponding to the
minimum value of the amplitude $A_{min}$. It is evident that the
distribution has maxima in the night-time and in the morning
hours, approximately from 00{:}00 to 07{:}00, and from 23{:}00 to
24{:}00 of local time LT.

Fig. $3{\bf d}$ presents the normalized probability distribution
$P(|A_{min}|)$ of TECS occurrence with a given amplitude
$A_{min}$. The vertical dashed line shows the threshold in
amplitude $\epsilon$ = 0.1 $TECU$. It was found that the largest
probable value of the amplitude $|A_{min}|$, also shown in
Fig.$3{\bf d}$, is about 0.3 $TECU$, and the half-width of the
distribution is 0.2 $TECU$. As demonstrated by Afraimovich et al.
(2001a), the mean values of the TEC variation amplitude with a
period of 20 min for the magnetically quiet and magnetically
disturbed days do not exceed 0.01 $TECU$ and 0.07 $TECU$,
respectively. Thus the most probable value of the amplitude
$A_{min}$ of the TECS identified here exceeds the mean values of
the TEC phase variation amplitude by a factor of 4-5 as a
minimum.

The availability of a large number of stations in some regions on
the globe, for instance, in California, USA, and West Europe,
furnishes an opportunity to determine not only the temporal but
also spatial characteristics of TECS. In order to estimate the
radius of spatial correlation of events of this type, the
number of cases was calculated where TECS within a single
2{.}3-hour time interval were observed at any two GPS stations
spaced by a distance $dR$. Fig. $3{\bf c}$ presents the histogram of the
number of such cases $P(dR)$ as a function of distance $dR$. It was
found that the localization of TECS in space is sharply defined.
In 90$\%$ of cases the distance $dR$ does not exceed
500 km.

\section{Dynamics and anisotropy of IIDs as deduced from the
October 5, 2001 event over California, USA}
\label{SPE-sect-4}

The event of October 5, 2001 was used in the analysis of the TECS
dynamic characteristics of TECS. On that day between 08:00 and
18:00 UT, a number of GPS stations located in California, USA
(220-260$^\circ$E; 28-42$^\circ$N) recorded a large number of
traveling ionospheric disturbances (TIDs) of the TECS type. For
the above-mentioned time interval and the selected longitude
range, the local time varied from 00:00 to 10:00 LT (for the
longitude of 240$^\circ$E corresponding to the center of the GPS
station array used in the analysis). So that the experimental
conditions were characteristic for the night-time ionosphere.

Fig. 4 illustrates the geometry of the experiment on October 5,
2001. Heavy dots show the GPS stations, and small dots indicate
the position of subionospheric points for all LOS's. Since several
(at least four) GPS satellites are observed at each receiving
site simultaneously, the number of LOS's far exceeds the number of
stations, which enhances considerably the possibilities of the
analysis. Panel {\bf a} presents the entire set of GPS stations
and subionospheric points that were used in the experiment for
the time interval from 8:00 to 10:00 UT. Panels {\bf b} and
{\bf c} show the stations and subionospheric points where the TEC
variations revealed TECS with an amplitude exceeding the
specified threshold $\epsilon$=0.01 $TECU$ ({\bf b}), and
$\epsilon$=0.1 $TECU$ ({\bf c}). Except in a single case, TECS
were recorded along the LOS running over land. As is evident from
the figure, the increase of the recording threshold did not
change the number of events recorded.

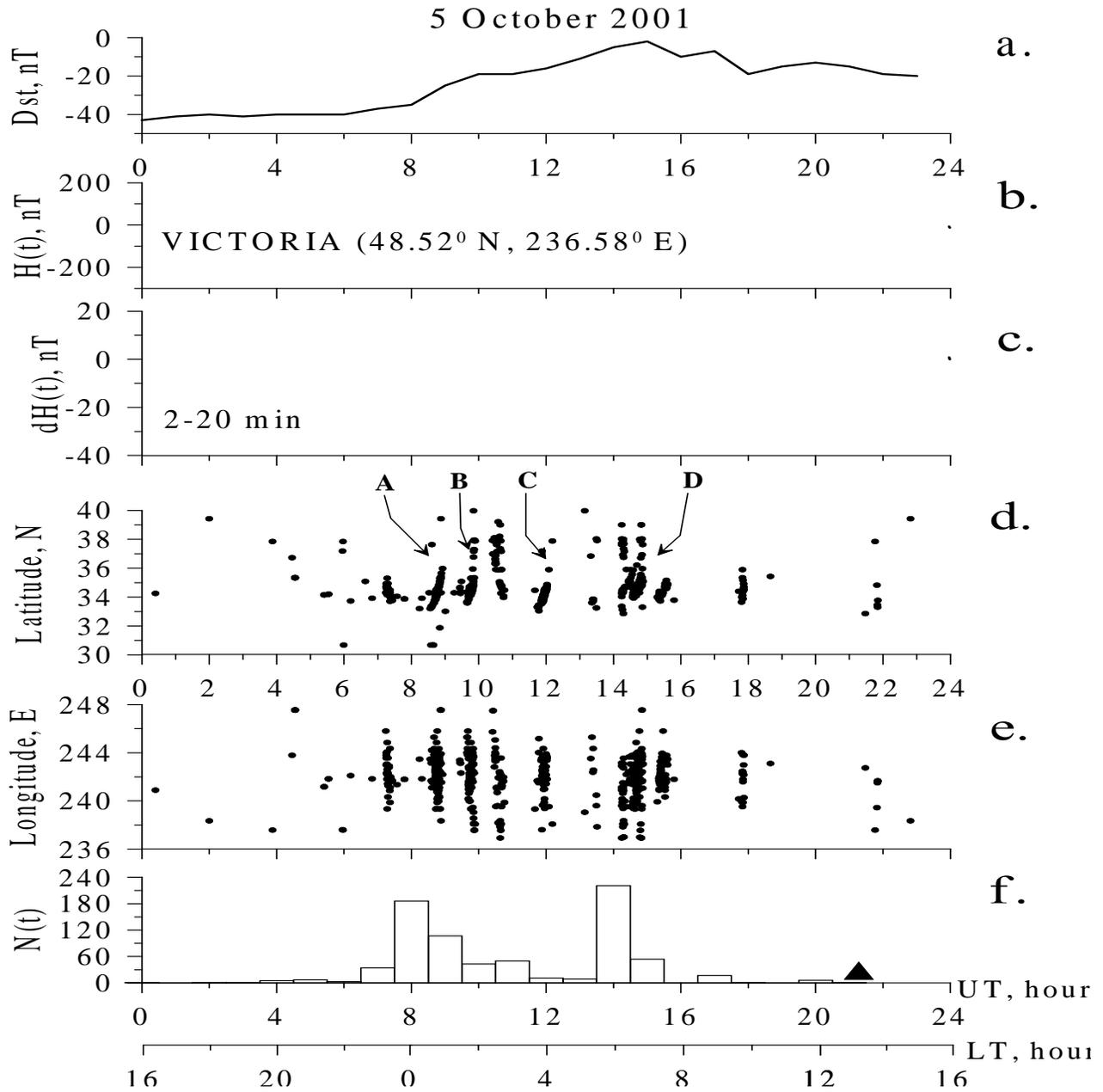
Fig. 5 presents the geomagnetic field $\it Dst$-variations ({\bf
a}) for October 5, 2001. $H(t)$-variations of the horizontal
component of the geomagnetic field as recorded at station
Victoria (48.52$^\circ$N; 236.58$^\circ$E) - {\bf b}.
$dH(t)$-variations of the horizontal component of the geomagnetic
field, filtered from the $H(t)$-series in the range of 2-20-min
periods - {\bf c}. At the lower time scale in Fig. 5, the local
time LT is represented for the longitude of 240$^\circ$E.

Fig. 5{\bf d} presents the distribution of the values of the GPS
station latitudes and time $t_{min}$, corresponding to each of
the TECS detected that day by all GPS stations of the California
region that were used in the analysis (220-260$^\circ$E;
28-42$^\circ$E). The letters A, B, C, and D in Fig. 5{\bf d}
label the TECS "traces" that are presented in Fig. 6 on a smaller
time scale (see Section 4.2). Fig. 5{\bf e} - same as in Fig.
5{\bf d}, but for the station longitudes and $t_{min}$.

Fig. 5{\bf f} presents the distribution $N(t)$ of the number of
TECS that were detected that day at all GPS stations used in the
analysis, with the rms above $\epsilon$=0.1 $TECU$.

\subsection{Determining the dynamic characteristics of IIDs by the
SADM-GPS method}
\label{SPE-sect-4.1}
The methods of determining the form and dynamic characteristics
of TIDs that are used in this study are based on those reported
in (Mercier, 1986; Afraimovich, 1997; Afraimovich et al.,
1998; 1999; 2000c).

We determine the velocity and direction of motion of the phase
interference pattern (phase front) in terms of some model of this
pattern, an adequate choice of which is of critical importance.
In the simplest form, space-time variations in phase of the
transionospheric radio signal that are proportional to TEC
variations $dI(t, x, y)$ in the ionosphere, at each given time $t$
can be represented in terms of the phase interference pattern that
moves without a change in its shape (the non dispersive
disturbances):

\begin{equation}
I(t,x,y)=F(t-x/u_x-y/u_y)
\label{IID-eq-03}
\end{equation}

where $u_x(t)$ and $u_y(t)$ are the displacement velocities of
intersection of the phase front of the axes x (directed to the
East) and y (directed to the North), respectively.

It should be noticed, however, that in real situations this
ideal model (3) is not realized
in a pure form. This is because that the TIDs
propagate in the atmosphere in the form of a dispersing wave
packet with a finite value of the width of the angular spectrum.
But in the first approximation on short time interval of
averaging compared to time period of filtered variations of TEC,
the phase interference pattern moves without a substantial
change in its shape.

A Statistical, Angle-of-arrival and Doppler Method (SADM) was
proposed by Afraimovich (1997) for determining the
characteristics of the dynamics of the phase interference pattern
in the horizontal plane by measuring variations of phase
derivatives with respect to the spatial coordinates $I'_x(t)$,
$I'_y(t)$, and to the time $I'_t(t)$. This permits the
determination of the unambiguous orientation of $\alpha(t)$ of
the wave-vector {\boldmath $K$} in the range 0--360${}^\circ$ and
the horizontal velocity $V_h(t)$ at each specific instant of
time.

Afraimovich et al. (1998, 1999, 2000c) described updating of the SADM
algorithm for GPS-arrays (SADM-GPS) based on a simple model for
the displacement of the phase interference pattern that travels
without a change in the shape and on using current
information about the angular coordinates of the LOS:
the elevation $\theta_s(t)$ and the azimuth $\alpha_s(t)$.

The method SADM-GPS makes it possible to determine the horizontal
velocity $V_h(t)$ and the azimuth $\alpha(t)$ of TID displacement
at each specific instant of time (the wave-vector orientation
{\boldmath $K$}) in a fixed coordinate system ($x$, $y$):

\begin{equation}
\begin{array}{rl}
\alpha(t) &=\arctan(u_y(t)/u_x(t))\\
u_x(t) &=I'_t(t)/I'_x(t) = u(t)/\cos\alpha(t) \\
u_y(t) &=I'_t(t)/I'_y(t) = u(t)/\sin\alpha(t) \\
u(t) &=|u_x(t) u_y(t)|/(u_x^2(t) + u_y^2(t))^{-1/2} \\
V_x(t) &=u(t)\sin\alpha(t)+w_x(t) \\
V_y(t) &=u(t)\cos\alpha(t)+w_y(t) \\
V_h(t) &= (V_x^2(t)+V_y^2(t))^{1/2}
\end{array}
\label{IID-eq-04}
\end{equation}

where $w_x$ and $w_y$ are the $x$ and $y$ projections of the
velocity $w$ of the subionospheric point (for taking into account
the motion of the GPS satellite).

Let us take a brief look at the sequence of data handling
procedures. Out of a large number of GPS stations, three
points (A, B, C) are chosen in such a way that the
distances between them do not exceed about one-half the expected
wavelength $\Lambda$ of the disturbance. The point B is
taken to be the center of a topocentric frame
of reference. Such a configuration of the GPS receivers
represents a GPS-array (or a GPS-interferometer) with a minimum
of the necessary number of elements. In regions with a dense
network of GPS-points, we can obtain a broad range of GPS-arrays
of a different configuration, which furnishing a means of testing
the data obtained for reliability; in this paper we have taken
advantage of this possibility (Section 4.2).

The input data include series of the vertical TEC $I_A(t)$,
$I_B(t)$, $I_C(t)$, as well as corresponding series of values of
the elevation $\theta_s(t)$ and the azimuth $\alpha_s(t)$ of the
LOS. Series of $\theta_s(t)$ and $\alpha_s(t)$ are used to
determine the location of the subionospheric point, as well as to
calculate the elevation $\theta$ of the wave vector {\boldmath
$K$} of the disturbance from the known azimuth $\alpha$ (see
formula (5)).

Since the distance between GPS-array elements (from several tens
to a few hundred of kilometers) is much smaller than that to
the GPS satellite (over 20000~km), the array geometry at the height
near the main maximum of the $F_2$-layer is identical to that on the
ground.

Linear transformations of the differences of the values of the
filtered TEC $(I_{{\rm B}}-I_{{\rm A}})$ and
$(I_{{\rm B}}-I_{{\rm C}})$ at the receiving
points A, B and C are used to calculate the components of the
TEC gradient $I'_x$ and $I'_y$ (Afraimovich et al., 1998).
The time derivative of TEC $I'_t$ is determined by differentiating
$I_{{\rm B}}(t)$ at the point B.

The resulting series are used to calculate instantaneous
values of the horizontal velocity $V_h(t)$ and the azimuth
$\alpha(t)$ of TID propagation. Next, the series $V_h(t)$ and
$\alpha(t)$ are put to a statistical treatment. This involves
constructing distributions of the horizontal velocity $P(V_h)$
and direction $P(\alpha)$ which are analyzed to test the hypothesis
of the existence of the preferred propagation direction. If such a
direction does exist, then the corresponding distributions are
used to calculate the mean value of the horizontal velocity
$\langle V_h\rangle$, as well as the mean value of the
azimuth $\langle \alpha\rangle$ of TID propagation.

The correspondence of space-time TEC characteristics, obtained
through transionospheric soundings, with local characteristics of
disturbances in the ionosphere was considered in detail in a wide
variety of publications (Afraimovich et al., 1992; Mercier and
Jacobson, 1997) and is not analyzed at length in this study. The
most important conclusion of the cited references is the fact
that, as for the extensively exploited model of a `plane phase
screen', disturbances $dI(x,y,t)$ of TEC faithfully copy the
horizontal part of the corresponding disturbance $dN(x,y,z,t)$ of
local concentration, independently of the angular position of the
source, and can be used in experiments on measuring the wave
velocity of TEC.

However, the TEC response amplitude experiences a strong `aspect'
dependence caused by the integral character of a transionospheric
sounding. As a first approximation, the transionospheric sounding
method is responsive only to TIDs with the wave vector {\boldmath
$K$} perpendicular to the direction {\boldmath $r$} of the
LOS. A corresponding condition for elevation $\theta$ and azimuth
$\alpha$ of {\boldmath $K$} has the form

\begin{equation}
\tan\theta=-\cos(\alpha_s-\alpha)/\tan\theta_s
\label{IID-eq-05}
\end{equation}

We used formula~(5) in determining the elevation
$\theta$ of {\boldmath $K$} from the known mean value of azimuth
$\alpha$ by Afraimovich et al. (1998).

The phase velocity modulus $V$ can be defined as

\begin{equation}
\label{IID-eq-06}
V=V_h\times cos(\theta)
\end{equation}

Hence at least the dynamic (two-dimensional) characteristics of
TEC disturbances (TECS in the case under consideration) may well
be referred qualitatively to the corresponding characteristics
$dN/N$ of local electron density disturbances (IIDs).
Reconstructing the quantitative characteristics of local density
disturbances, $dN/N$, in terms of the solution of an inverse
problem of transionospheric sounding is a highly difficult,
special problem, constituting the subject of our further
investigation.

On the basis of using the transformations described in this
section, for each of the GPS arrays chosen for the analysis we
obtained the average for the selected time interval values of the
following TECS parameters: $\langle \alpha\rangle$ and $\langle
\theta\rangle$ -- the azimuth and elevation of the wave vector
{\boldmath $K$}; $\langle V_h\rangle$ and $\langle V\rangle$ --
the horizontal component and the phase velocity modulus, the
azimuth of a normal to the TECS front $\alpha_c$ from the method
reported by Mercier (1986).

\subsection{The dynamic characteristics of IIDs}
\label{SPE-sect-4.2}

For the PRN05 and PRN30 satellites, Fig. 1{\bf c} and 1{\bf f}
give an example of the filtered TECS for different spaced
stations, and for the GPS satellites on October 5, 2001. It is
evident from the figure that the selected typical TEC variations
such as TECS are identical and shifted by a certain amount of
delay, which makes it possible to calculate the velocity and
direction of the IID that causes the observed TEC variations. The
panels show the values of the IID velocity $V$ and direction
$\alpha$ inferred using the above data processing procedures.

Using different sets of GPS arrays for the entire California
region we were able to obtain stable mean estimates of the IID
velocity modulus $V$, horizontal projection $V_h$ and direction
$\alpha$ in the horizontal plane, as well as the elevation angle
$\theta$ of the displacement vector {\boldmath $K$} in the
vertical plane.

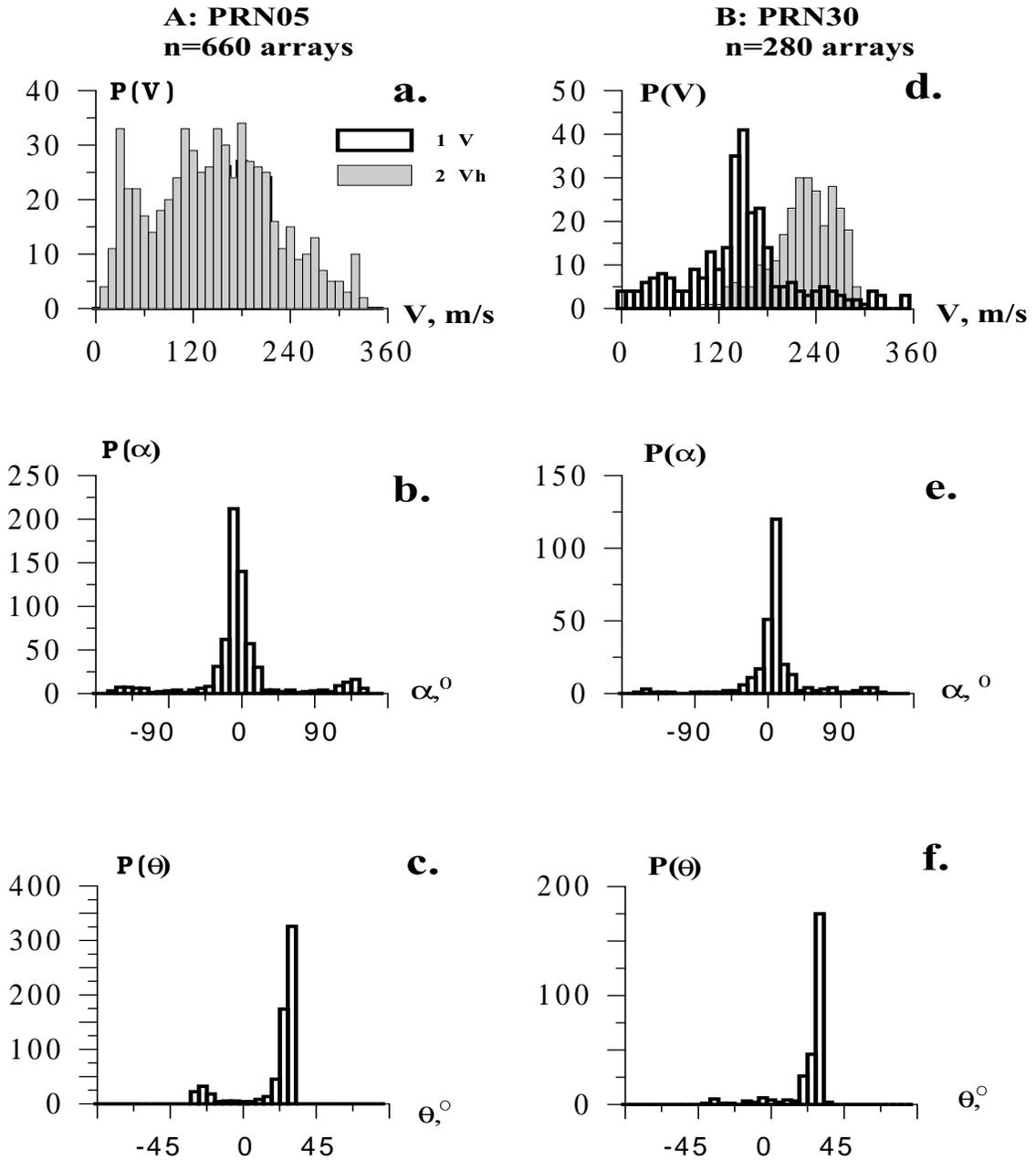
Fig. 7. present the distributions of the TECS parameters as
determined by the SADM-GPS method, for the "trace" A (on the
left; 660 arrays), and for the "trace" B (on the right; 280
arrays). {\bf a}, {\bf d} -- modulus $V$ (line 1) and horizontal
component $V_h$ (line 2) of the TECS phase velocity; {\bf b},
{\bf e} -- azimuth $\alpha$; {\bf c}, {\bf f} -- elevation
$\theta$ of the TECS wave vector {\boldmath $K$}. Same parameters
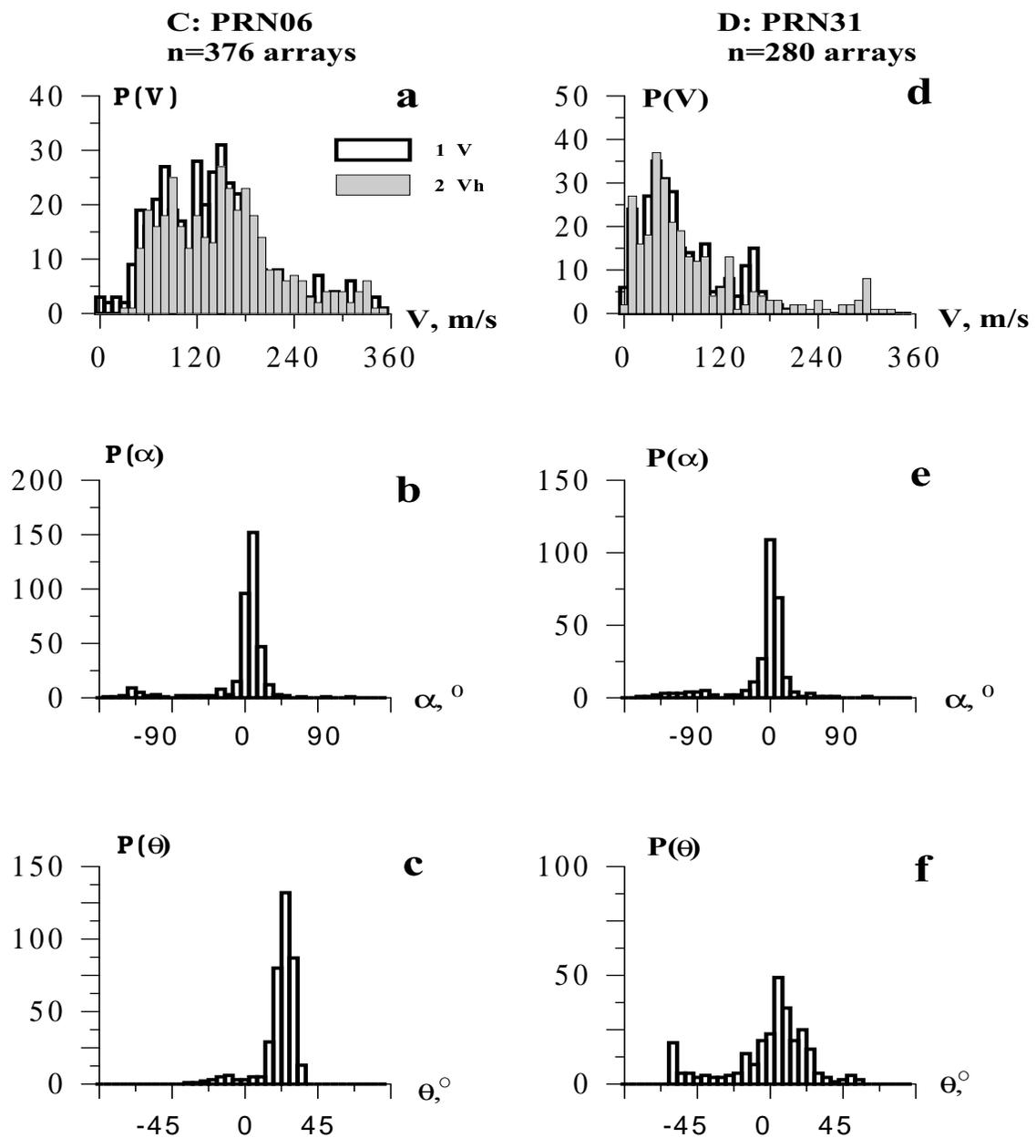
as in Fig.7, but for the "trace" C (on the left; 376 arrays), and
for the "trace" D (on the right; 280 arrays), are presented on
Fig.8.

According to our data, the following values were obtained (using
SADM-GPS method):
$<V_h>$=179 m/s, $V$=160 m/s, $\alpha$=360$^\circ$, and
$<\theta>$=22$^\circ$ for the trace" A (Fig.7, on the left);
$<V_h>$=190 m/s, $V$=165 m/s, $\alpha$=5$^\circ$, and
$<\theta>$=27.6$^\circ$ for the trace" B (Fig.7, on the right);
$<V_h>$=171 m/s, $V$=151 m/s, $\alpha$=5$^\circ$, and
$<\theta>$=24$^\circ$ for the trace" C (Fig.8, on the left); and
$<V_h>$=90 m/s, $V$=74 m/s, $\alpha$=360$^\circ$, and
$<\theta>$=20$^\circ$ for the trace" D (Fig.8, on the right).

An analysis of the distribution of the azimuths $P(\alpha)$
(Figs.~7b, 7e, 8b, and 8e) shows a clearly pronounced northward direction
of TECS displacement. The elevation of the TECS wave vector,
determined from the aspect condition (5), has mostly a small
positive value (Figs.~7c, 7f, 8c, and 8f).

\subsection{The anisotropy and sizes of IIDs}
\label{SPE-sect-4.3}

Let us consider in greater detail the "traces" A, B, C, and D.
Fig. 6 (a - d) plots the dependencies of the values of the GPS
station latitudes $N_i$ on the time $t_{min, i}$, corresponding
to each of the TECS detected that day at all GPS stations of the
California region used in the analysis (220-260$^\circ$E;
28-42$^\circ$N); fig. 6 (e - h) - same as above, but for the
station longitudes $E_i$ and $t_{min, i}$. TECS only for a single
selected satellite number are considered for each trace.

Figs.6 {\bf (a - d)} suggest the conclusion that at each given time
there are only a few TECS with close values of the coordinates
$N_i$, and there is a clearly pronounced gradual displacement of the
subionospheric point in the northward meridional direction. With
a knowledge of the range of TECS displacement in latitude and the
time interval corresponding to this displacement, it is easy to
determine that the meridional projection of the displacement
velocity $V_N$ is close to 200 m/s, which is in agreement with
estimates obtained using the SADM-GPS method.

On the other hand, the dependencies of the longitude $E_i$ of the
TECS observed on $t_{min, i}$ suggest a spatial extent of the
traveling irregularity in longitude.

Thus the shape of the observed irregularities
("traces" A, B, C, and D) show up as ellipsoids traveling
northward with a velocity of about 200 m/s.

The IID anisotropy was analyzed by determining the "contrast" $C$
(Mercier, 1986). We calculated the ratio $C_{N,E}$:

\begin{equation}
\begin{array}{rlrl}
C_{N,E}&=\sigma_N/\sigma_E, & \mbox{if } \sigma_N&>\sigma_E\\
C_{N,E}&=\sigma_E/\sigma_N, & \mbox{if } \sigma_E&>\sigma_N
\end{array}
\label{IID-eq-07}
\end{equation}

where $\sigma_N$ and $\sigma_E$ are the standard deviations of
the corresponding series of the TECS coordinates $N_i$ and $E_i$.
These series were obtained by transforming the initial series
$N'_i$ and $E'_i$ by rotating the original coordinate system (N,
E) by an angle $\beta$:

\begin{equation}
\begin{array}{rl}
N&=N'\sin\beta+ E'\cos\beta\\
E&=-N'\cos\beta+E'\sin\beta \end{array}
\label{IID-eq08}
\end{equation}

Mercier (1986) showed that it is possible to find such a value
of the rotation angle $\beta_0$, at which the ratio $C_{N,E}$
will be a maximum and equal to the value of contrast $C$. This
parameter characterizes the degree of anisotropy of the phase
interference pattern. The angle $\beta_0$ in this case
indicates the direction of elongation, and the angle
$\alpha_c=\beta_0+\pi/2$ indicates the direction of the wave
vector {\boldmath $K$} coincident (module 180${}^\circ$) with
the propagation direction of the phase front.

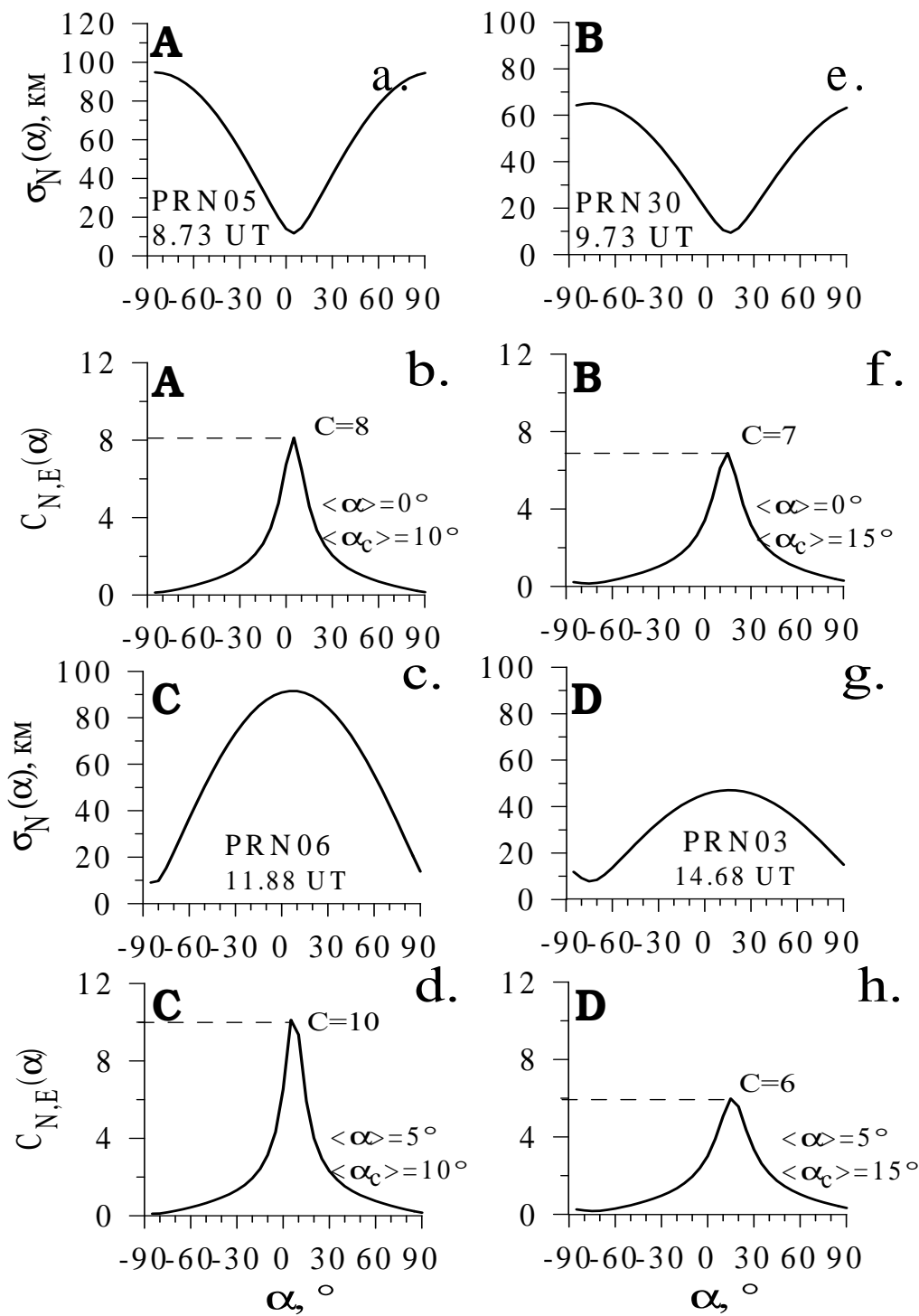
Fig. 9 ($\bf a, c, e, g$) plots the dependencies of the standard
deviation $\sigma_N(\alpha)$ of the values of the coordinates of
subionospheric points in a topocentric coordinate system, and of
the corresponding to $t_{min}$, on the rotation angle
$\alpha_c$=$\beta_0$+$\pi$/2 ($\beta$ stands for the rotation
angle of the original coordinate system) for the "traces" A, B,
C, and D (Figs. 5 and 6); Fig. 9 ($\bf b, d, f, h$) plots the
dependencies of the value of the ratio $C_{N,E}$ on
$\alpha_c$=$\beta_0$+$\pi$/2. The values of the contrast $C$ are
shown by the horizontal dashed line.

It is apparent from the figure that
the value of $C$ fluctuates from 6 to 10, and the greatest contrast
for the traces under consideration corresponds to the "trace" C
and, hence, this trace has the greatest anisotropy.

Furthermore, the direction of elongation of the major axis of the
IID ellipse $\beta_0$ = -85$^\circ$. For the other "traces" the
values of the parameter $\beta_0$ lie in the range from
-80$^\circ$ to -85$^\circ$. The panels show the mean values of
the angles $\alpha_c$ obtained by the method reported in Mercier
(1986), and the mean values of $\alpha$ inferred by the SADM-GPS method
(Afraimovich, 1997; Afraimovich et al., 1998; 1999; 2000c) for
each of the "traces". As is evident from the figure, the values
of the angle $\alpha$, calculated by the two methods, agree to
within 5-10$^\circ$.

We now estimate the horizontal sizes of the IID. With the mean
duration of 10 min and the travel velocity of 180 m/s, we obtain
the transverse size of the irregularity equal to 108 km. With the
value of the contrast $C$=9, the longitudinal size equal to 864
km.

\subsection{Estimating the relative IID amplitude}
\label{SPE-sect-4.4}

Let us now obtain the mean estimate of the relative amplitude of
a local electron density disturbance typified by the IID of
October 5, 2001.

A mean TECS absolute amplitude over California equal 0.1 $TECU$
(see Fig.1{\bf c, f}). As the background value of $I_0$, we used
the absolute "vertical" TEC value of $I_0(t)$ obtained from
IONEX-maps of the TEC (see Fig.10{\bf c}). These maps with
two-hour temporal resolution were constructed using the
well-known methods and are placed on the Internet site
ftp://cddisa.gsfc.naa.gov/.

Fig.10{\bf c} shows the variations of the absolute "vertical" TEC
value $I_0(t)$ for the site with the coordinates 240${}^\circ$E;
35${}^\circ$N. At the lower time scale in Fig.10, the local time
LT is represented for the longitude of 240$^\circ$E. For
comparison Fig.10{\bf d} present the distribution $N(t)$ of the
number of TECS that were detected that day at all GPS stations
used in the analysis, with the rms above $\epsilon$=0.1 $TECU$
(this is copy of Fig.5{\bf f}).

Comparison of Fig. 10{\bf c} and {\bf d} reveals that the
greatest TECS occurrence probability corresponds to the
night-time hours for which the "vertical" TEC value $I_0(t)$ does
not exceed 10 $TECU$.

Hence the relative amplitude of TECS $dI/I_0$ makes up 1$\%$,
that is quite significant for this disturbance period ($\Delta T$
= 10 min) and exceeds one order of magnitude the amplitude of
typical background TEC fluctuations (Afraimovich et al., 2001a).

It is further assumed that the characteristic vertical size of
the IID is of the same order as the transverse horizontal size
(of about 100 km - see Section 4.3). The vertical extent of the
part of the ionosphere that makes the main contribution to the
TEC modulation is no less than 500-1000 km. Hence it follows that
the relative amplitude $dN/N$ of the local electron density
disturbance for IID reaches a considerably large value,
5--10$\%$.

\section{Discussion}
\label{SPE-sect-5}

What is the nature of the ionospheric irregularities that are
responsible for the occurrence of TECS, and Do they differ from
the known published ionospheric disturbances? We shall try to
unravel this situation using the sample statistic obtained in
Section 3 and the October 5, 2001 event as an example, because
for this event it was possible not only to record a large number
of TECS but also to obtain estimates of the size, anisotropy and
velocity of isolated ionospheric irregularities that are
responsible for the occurrence of TECS (Section 4).

It is significant that, according to Internet data for the
concerned region of the USA and for the time interval of October
5, 2001 of our interest, there were not observable meteorological
phenomena (site http://www.solar.ifa.hawaii.edu/Tropical/),
powerful explosions and rocket launchings able
to cause TEC variations of the TECS type. The $Titan-4$ rocket
was launched from the "Vandenberg" spaceport ($239.5^\circ$E;
$34.8^\circ$N) on October 5 at 21:21 UT (the launching time is
shown by the filled triangle in Fig.5{\bf f}). However, most TECS
on that day were observed before the launching time.

As far as seismic activity is concerned, we analyzed the data
covering not only the time interval of our interest but also for
the period from October 4 to 20, 2001
(http://wwwneic.cr.usgs.gov/neis/FM/previous/0107.html). No
earthquakes with the magnitude larger than 3.5 were recorded in
the region of our interest for the above time interval.

For the October 5, 2001, time interval of our interest, TEC
fluctuations similar to TECS detected over California were absent
elsewhere on the globe. No solar flare-induced background
fluctuations of the TEC disturbance were also revealed. A
similar, quiet, behavior over this time interval was also
characteristic for the energetic particle flux. We do not present
here the relevant data for reasons of space. They may be found on
the site http://www.sel.noaa.gov/ftpmenu/lists.html.

The geomagnetic situation on October 5, 2001 can be characterized
as a weakly disturbed one, which must lead to some increase of
the level of TEC background fluctuations; however, this cannot
cause any large-scale changes in electron density which are
characteristic for the geomagnetically disturbed ionosphere. In
the analysis of the geomagnetic situation we used the data from
magnetic observatory Victoria (48.52$^\circ$N; 236.58$^\circ$E)
which, for the time interval 15:00 - 18:00 UT of our interest,
recorded a geomagnetic disturbance implying a decrease of the
horizontal component of the magnetic field by 100 nT (Fig. 5{\bf
b}).

This disturbance was also accompanied by an increase of the
$H$-component fluctuation amplitude in the range of 2-20-min
periods (Fig.5{\bf c}). The variation range of the geomagnetic
$\it Dst$-index for the selected time interval was also
relatively small (no more than 20 nT); however, the period from
08:00 to 15:00 UT showed a clearly pronounced increase in
variations of the $\it Dst$-index that coincided with the period
when the $H$-component of the magnetic field was increasing (Fig.
5{\bf a}).

Hence, the data from magnetic-variation stations do not suggest
the conclusion that the observed TECSs are associated with
magnetic field variations.

\subsection{The difference of TECS from the TEC response to
shock-acoustic waves generated due to rocket launchings and
earthquakes}
\label{SPE-sect-5.1}

First of all, it is important to remark that TECS
variations are close in amplitude, form and duration to the TEC
response to shock-acoustic waves (SAW) generated during rocket
launchings (Calais and Minster, 1996; Calais et al., 1998b; Li et
al., 1994; Afraimovich et al., 2000a; 2001b), earthquakes (Calais
and Minster, 1995; Afraimovich et al., 2001c), and explosions
(Fitzgerald, 1997; Calais et al., 1998a). In this connection,
TECS can mask TEC responses to technogenic effects and lead to
spurious signals recorded in detection systems for such effects,
based on analyzing signals from the global GPS network.

Afraimovich et al. (2000a; 2001b; 2001c) found that spite of a
difference of rockets and earthquakes characteristics, the
ionospheric TEC response for all such events had the character of
an $N$--wave corresponding to the form of a shock wave. The SAW
period $\Delta T$ is 270--360~s, and the amplitude exceeds the
standard deviation of total electron content background
fluctuations in this range of periods under quiet and moderate
geomagnetic conditions by factors of 2 to 5 as a minimum. The
angle of elevation $\theta$ of the SAW wave vector varies from
30${}^\circ$ to 60${}^\circ$, and the SAW phase velocity $V$
(900--1200~m/s) approaches the sound velocity at heights of the
ionospheric $F$-region maximum.

A spatial and temporal processing of data from GPS arrays can be
used in selecting these phenomena in order to estimate the
propagation velocity of TEC disturbances. In particular, the
velocity of IIDs (150--200 m/s) is far less than the SAW velocity
(900--1200~m/s) and is most likely to correspond to the
propagation velocity of medium-scale background AGW (Kalikhman,
1980; Afraimovich et al., 1998; 1999; Mercier, 1986; Mercier and
Jacobson, 1997; Hocke and Schlegel, 1996; Oliver et al., 1997).
The SAW and IID responses can be distinguished using this
attribute, as well as from the difference of angular
characteristics of the disturbance wave vector (see Section 4).

\subsection{Can TECS be caused by the $E_s$-layer ionospheric
irregularities that are responsible for S- and
QP-scintillations?}
\label{SPE-sect-5.2}

As has been pointed out in the Introduction, a large number of
publications (e.g., Karasawa, 1985; Titheridge, 1971; etc.), were
devoted to the study the "spikes-type" (S-) scintillations.

Bowman (1989) observed quasi-periodic (QP) oscillations of
the signal from an orbiting satellite at 150 MHz frequency. The
anticipated cause of the occurrence of such QP oscillations is
Fresnel diffraction from ionospheric irregularities. Evidently,
irregularities giving rise to QP oscillations reside in the
$E_s$ level containing a host of small regions (less than 200
m in size) with high electron density separated from one another
by a distance of several tens of kilometers. The occurrence of QP
oscillations is peaked in the night-time hours. The calculated
velocities of irregularities and $E_s$ structures range from
50 to 100 m/s.

The cited authors point out that transionospheric signal
variations of this kind are observed mostly at night and are most
likely caused by the irregularities located in the $E_s$-layer.
This brings up the question of whether the TECS recorded by us
are able to be caused by such ionospheric irregularities. At this
point it should be noted that TECS are also observed mostly at
night, and their velocities (on the order of 100 km/s) are close
to those obtained by Karasawa (1985), Titheridge (1971), and Bowman
(1989).

In order to verify the validity of this TECS model for the
October 5, 2001 event we availed ourselves of the data (placed on
the site http://spidr.ngdc.noaa.gov/spidr/) from the ionospheric
station Point Arguello located in the center of the region of our
interest.

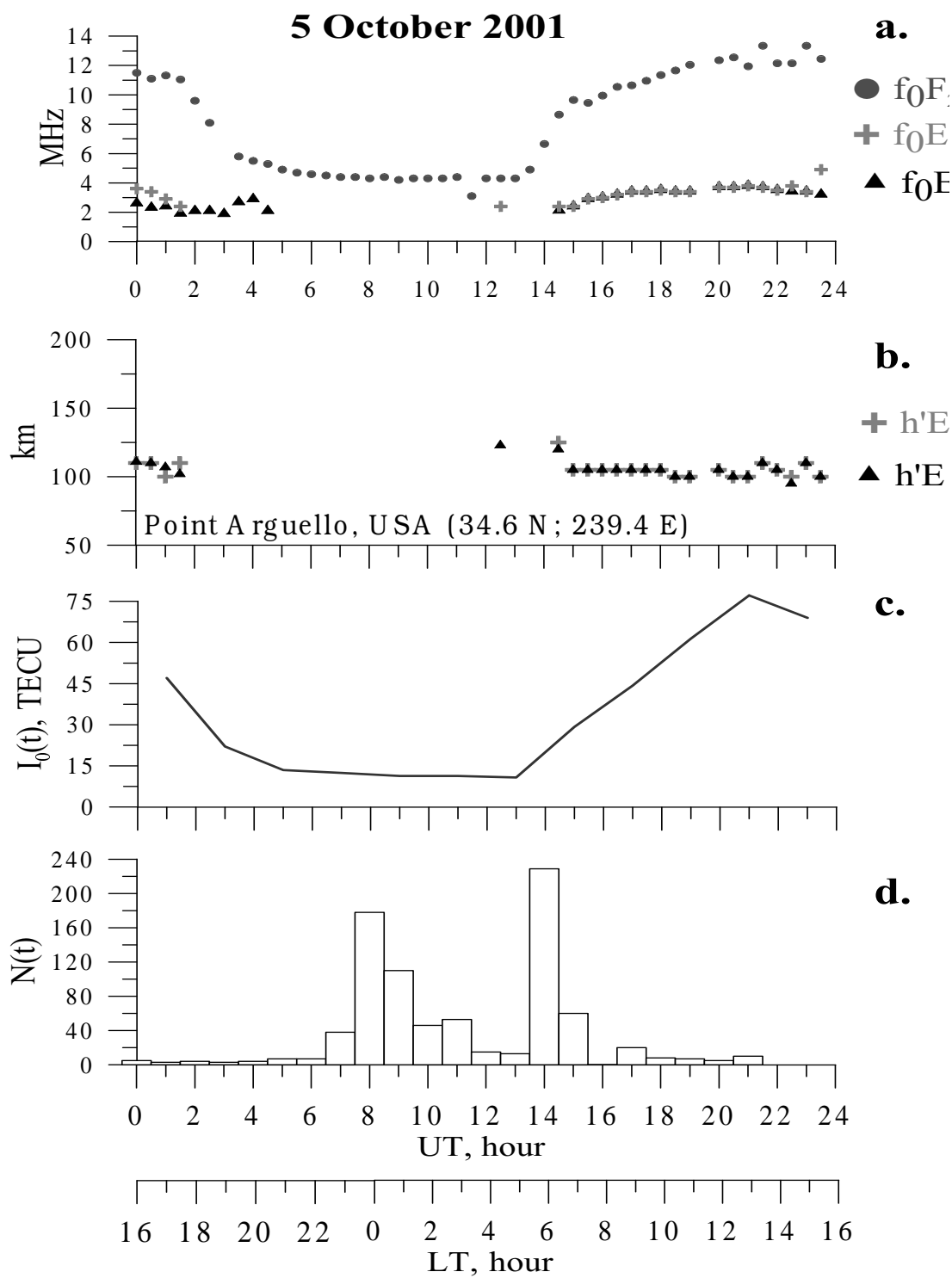
Fig.10{\bf a} present the values of the critical frequencies
$f_{0}F_2$ -- heavy dots; $f_{0}E_s$ -- crosses; $f_{0}E$
--shaded triangles, measured on October, 5, 2001 at ionospheric
station Point Arguello (239.4${}^\circ E$; 34.6${}^\circ N$);
Fig.10{\bf b} -- variations of virtual height $h'E$ -- shaded
triangles, $h'E_s$ -- crosses.

An analysis of the dependence of critical frequencies and
effective heights of reflections from the $F$- and $E_s$-regions
showed that during 08-14 UT when the largest number of TECS was
recorded, reflections from the $E_s$-layer were absent altogether
(cf. panels {\bf a, b, d}).

Thus we did not obtain any direct confirmation of the fact that
TECS were caused by intense ionospheric irregularities located in
the $E_s$-layer. The question of the origin of the IID that are
responsible for TECS remains open.

It should be noted that direct comparison of our results with the
data obtained by Karasawa (1985), Titheridge (1971), and Bowman
(1989), is made difficult by the fact that the overwhelming
amount of earlier data was obtained for amplitude, rather than
phase, scintillations. On the other hand, extracting data on
amplitude variations of GPS signals from Internet RINEX-files is
highly difficult (Gurtner, 1993).

To understand the nature of TECS requires invoking data from
different, independent diagnostic tools, including incoherent
scatter radars, ionosondes, magnetometers, etc.

\section{Conclusion}
\label{SPE-sect-6}

Main results of this study may be summarized as follows:
\begin{enumerate}

\item TECS constitute a rare event that occurs mainly in the
night-time in the spring and autumn periods, in a weakly
disturbed or quiet geomagnetic situation.

\item In the time region, TECS represent single negative
aperiodic abrupt changes of TEC with a duration of about 10-20
min; the mean value of the TECS amplitude exceeds the mean value
of the TEC variation amplitude by a factor of 4-5 as a minimum
and is 0.2 $TECU$.

\item TECS represent a local phenomenon with a typical radius of
spatial correlation not larger than 500 km.

\item The TECS we recorded on October 5, 2001, were caused by
isolated ionospheric irregularities of the ellipsoid shape, with
the direction of elongation of the major axis of the ellipse on
the order of -85$^\circ$. IIDs travel in a direction
perpendicular to their elongation (northward in the case under
consideration) with the mean velocity $<V>$=160 m/s, which
corresponds to the velocity of medium-scale AGW (of about
100-200 m/s). The size of the observed irregularities is 100 km
by 800 km, respectively.
\end{enumerate}

We are aware that this study has revealed only the key averaged
patterns of this phenomena, and we hope that it would give
impetus to a wide variety of more detailed investigations.

\section*{ACKNOWLEDGMENTS}

We are indebted to Dr. N.T. Afanasyev for participating in
discussions. We thank O.S. Lesyuta and S.V. Voeykov for help in
organizing the experiment. We are also grateful to V.G.
Mikhalkovsky for his assistance in preparing the English version
of the manuscript. This work was done with support under RFBR
grant of leading scientific schools of the Russian Federation No.
00-15-98509 as well as Russian Foundation for Basic Research
grants No. 00-05-72026 and 02-05-64570.

\end{document}